\documentclass[numberedappendix,letterpaper]{emulateapj}
\usepackage{mathptmx}

\def\vi{$V_{606}-I_{814}$\/ }
\def\dvi{$\delta(V-I)$\/ }
\def\ltsima{$\; \buildrel < \over \sim \;$}
\def\simlt{\lower.5ex\hbox{\ltsima}}
\def\gtsima{$\; \buildrel > \over \sim \;$}
\def\simgt{\lower.5ex\hbox{\gtsima}}
\def\simless{\mathbin{\lower 3pt\hbox
   {$\rlap{\raise 5pt\hbox{$\char'074$}}\mathchar"7218$}}}   
\def\simgreat{\mathbin{\lower 3pt\hbox
   {$\rlap{\raise 5pt\hbox{$\char'076$}}\mathchar"7218$}}}   

\def\ib{$I_{814}$\/ }

\def\vb{$V_{606}$\/ }

\def\vi{$V_{606}-I_{814}$\/ }
\def\bv{$g_{450}-V_{606}$\/ }

\def\dvi{\mbox{$\delta(V-I)$}}

\def\vir{\mbox{$(V_{606}-I_{814})(r)$}}

\slugcomment{Accepted for publication in the Astrophysical Journal}

\shorttitle{Resolving spheroid galaxies from HST/ACS WFC UGC10214 ERO observations}
\shortauthors{F. Menanteau \& The ACS Science Team}

\begin{document}

\title{Internal Color Properties of Resolved Spheroids in the Deep HST/ACS field of UGC 10214}

\author{F. Menanteau\altaffilmark{1},
H.C. Ford\altaffilmark{1},
G.D. Illingworth\altaffilmark{3},
M. Sirianni\altaffilmark{1}, 
J.P. Blakeslee\altaffilmark{1},
G.R. Meurer\altaffilmark{1},
A.R. Martel\altaffilmark{1},
N. Ben\'{\i}tez\altaffilmark{1},
M. Postman\altaffilmark{2},
M. Franx\altaffilmark{7},
D.R. Ardila\altaffilmark{1}
F. Bartko\altaffilmark{4}, 
R.J. Bouwens\altaffilmark{1},
T.J. Broadhurst\altaffilmark{5},
R.A. Brown\altaffilmark{2},
C.J. Burrows\altaffilmark{2},
E.S. Cheng\altaffilmark{6},
M. Clampin\altaffilmark{2},
N.J.G. Cross\altaffilmark{1},
P.D. Feldman\altaffilmark{1},
D.A. Golimowski\altaffilmark{1},
C. Gronwall\altaffilmark{8},
G.F. Hartig\altaffilmark{2},
L. Infante\altaffilmark{9}
R.A. Kimble\altaffilmark{6},
J.E. Krist\altaffilmark{2},
M.P. Lesser\altaffilmark{10},
G.K. Miley\altaffilmark{7},
P. Rosati\altaffilmark{11}, 
W.B. Sparks\altaffilmark{2}, 
H.D. Tran\altaffilmark{1}, 
Z.I. Tsvetanov\altaffilmark{1},   
R.L. White\altaffilmark{1,2}
\& W. Zheng\altaffilmark{1}}


\altaffiltext{1}{Department of Physics and Astronomy, Johns Hopkins
University, 3400 North Charles Street, Baltimore, MD 21218.}

\altaffiltext{2}{STScI, 3700 San Martin Drive, Baltimore, MD 21218.}

\altaffiltext{3}{UCO/Lick Observatory, University of California, Santa
Cruz, CA 95064.}


\altaffiltext{4}{Bartko Science \& Technology, P.O. Box 670, Mead, CO
80542-0670.}


\altaffiltext{5}{Racah Institute of Physics, The Hebrew University,
Jerusalem, Israel 91904.}


\altaffiltext{6}{NASA Goddard Space Flight Center, Laboratory for
Astronomy and Solar Physics, Greenbelt, MD 20771.}


\altaffiltext{7}{Leiden Observatory, Postbus 9513, 2300 RA Leiden,
Netherlands.}


\altaffiltext{8}{Department of Astronomy and Astrophysics, The
Pennsylvania State University, 525 Davey Lab, University Park, PA
16802.}


\altaffiltext{9}{Departmento de Astronom\'{\i}a y Astrof\'{\i}sica,
Pontificia Universidad Cat\'{\o}lica de Chile, Casilla 306, Santiago
22, Chile.}


\altaffiltext{10}{Steward Observatory, University of Arizona, Tucson,
AZ 85721.}


\altaffiltext{11}{European Southern Observatory,
Karl-Schwarzschild-Strasse 2, D-85748 Garching, Germany.}

\begin{abstract}

We study the internal color properties of a morphologically selected
sample of spheroidal galaxies taken from the Hubble Space Telescope
(HST) Advanced Camera for Surveys (ACS) ERO program of UGC 10214
(``The Tadpole''). By taking advantage of the unprecedented high
resolution of the ACS in this very deep dataset we are able to
characterize spheroids at sub-arcseconds scales. Using the \vb and \ib
bands, we construct $V-I$ color maps and extract color gradients for a
sample of spheroids at $I_{814W} < 24$ mag. We assess the ability of
ACS to make resolved color studies of galaxies by comparing it with
the multicolor data from the Hubble Deep Fields (HDFs). Here, we
report that with ACS/WFC data of $\simgt 10\times$ less exposure time
than in the WFPC2 HDFs it is possible to confidently carry out
resolved studies of faint galaxies at similar magnitude limits. We
also investigate the existence of a population of morphologically
classified spheroids which show extreme variation in their internal
color properties similar to the ones reported in the HDFs. These are
displayed as blue cores and inverse color gradients with respect to
those accounted from metallicity variations. Following the same
analysis we find a similar fraction of early-type systems
($\sim30\%-40\%$) that show non-homologous internal colors, suggestive
of recent star formation activity. We present two statistics to
quantify the internal color variation in galaxies and for tracing blue
cores, from which we estimate the fraction of non-homogeneous to
homogeneous internal colors as a function of redshift up to $z
\simless 1.2$. We find that it can be described as about constant as a
function of redshift, with a small increase with redshift for the
fraction of spheroids that present strong color dispersions. The
implications of a constant fraction at all redshifts suggests the
existence of a relatively permanent population of evolving spheroids
up to $z\simless 1$. We discuss the implications of this in the
context of spheroidal formation.

\end{abstract}


\section{Introduction}

In recent years, the use of field elliptical galaxies has become a
conventional tool for testing between very different structures of
formation, where the main competing views of galaxy formation at high
and low redshift are often referred as monolithic or early formation
and hierarchical or late formation \citep{Peebles-02}. Many
observational studies have been devoted to study a number of ``scaling
relations'' in ellipticals in rich clusters; i.e. the low scatter in
the fundamental plane \citep{Djorgovski-Davis-87,
Dressler-etal-87,vanDokkum-etal-96, Treu-etal-1999} and in the
color-magnitude relation (CMR) \citep{Sandage-etal-78, BLE92,
Ellis-etal-97}. These imply a high degree of homogeneity in the
stellar population which reinforced the idea of a monolithic collapse
model in which ellipticals formed during a rapid
collapse at high redshift \citep{Eggen-etal-62}.

The simple view of early formation contrasts with the predictions of
models where galaxies assemble hierarchically as the result of the
merging of smaller sub-units, at a rate governed by the merger of cold
dark matter halos \citep{White-Rees-78, White-Frenk-91,
Baugh-etal-96}. Although the prevailing view in the past was that
ellipticals in the field formed in isolation in a high-redshift
initial burst of formation, as did their clustered counterparts,
several authors have shown that the observational properties of field
spheroids (collectively E and S0s) are incompatible with the bulk of
the population forming at high redshift \citep{Zepf-97,
Barger-etal-99, Menanteau-etal-99, Trager-etal-2000}.
Most of these studies were prompted by the success of hierarchical
models in the predictions of a large quantity of observable properties
of galaxies, suggesting that massive field ellipticals could only have
been assembled recently ($z<1$). The general consensus that arose is
that a single short period of formation at high redshift is
incompatible with the observations. There is growing evidence of the
existence of blue spheroids with perturbed colors at intermediate
redshift and a deficit of red systems compared to monolithic models
predictions \citep{Treu-Stiavelli-99,
Menanteau-etal-01,KBB99,Zepf-97}. More recently, however,
\cite{Bell-etal-2004} have reported using a large sample of HST
galaxies that E/S0s dominate the red-sequence of galaxies at
$z\sim0.7$. As the outcome of these new studies a new view of
elliptical formation is emerging.

A surprising result from this wave of studies is the existence of
field spheroids with blue central regions at intermediate redshifts
$0.4<z<1.0$. These objects were originally detected from their \vi
color maps (\citealt{Abraham-etal-99}; \citealt*{Menanteau-etal-01})
in the {\em Hubble Deep Fields} (HDFs), and are suggestive of recent
star-formation activity associated to central region of the galaxy.
The presence of blue core spheroids in the HDFs has generated
interest in reproducing their observed properties using both
semi-analytic models \citep*{Benson-Ellis-Menanteau-02} and extended
monolithic collapse (\citealt{Jimenez-etal-98};
\citealt*{Menanteau-Jimenez-01}). The hierarchical description of
\cite*{Benson-Ellis-Menanteau-02} accurately predicts the number of
spheroids and the degree of color variations observed in spheroids,
but misses most of the red spheroids in the sample. On the other hand
multi-zone chemo-dynamical models can predict the existence of blue
cores and homogeneous colors of evolved ellipticals with success, by
adjusting the redshift of formation of the galaxy, although the
cosmological mechanisms responsible for detailing the formation of
spheroids are somewhat
unclear. \citep*{Menanteau-Jimenez-01,Friaca-Terlevich-01}

The use of resolved colors in evolutionary studies, although largely
unexploited for characterizing distant galaxies \citep[see][for
examples]{Tamura-etal-00,Abraham-etal-99,Menanteau-etal-01,Hinkley-Im-01}
supplies a new set of independent constraints to the traditional use
of counts and redshift distributions alone, particularly useful when
attempting to discriminate between models of elliptical
formation. Moreover, it is appealing to examine independent spheroid
samples that look for galaxies with color inhomogeneities like those
reported in the HDFs. Over larger datasets, they can provide important
clues for determining the fraction of spheroids still experiencing
star formation. Unfortunately, until the arrival of the ACS, resolved
colors analysis have been limited only to superb signal-to-noise and
highly time consuming datasets such as the HDFs. In this paper we
explore the advantages arising from exploiting the unprecedented
capabilities of the ACS in obtaining HDF-like datasets with only a
fraction of the integration time.  We use the first data available
from the ACS to study the resolved color properties of faint distant
galaxies and compare them with previous results from the HDFs.

A plan for the paper follows. In Section 2, we describe the
observations and data reduction of the ACS images and our
morphological classification of galaxies utilized in segregating
spheroids. In Section 3 we discuss the resolved color properties and
the construction of color maps of spheroids in our sample. In Section
4 we present our methodology for characterizing the resolved color
properties using a model-independent approach. In Section 5, we
discuss the results of our analysis and finally in section 6 we
summarize our conclusions.

\section{Spheroids in the ACS/WFC UGC 10214 field}

\begin{deluxetable*}{lrrccccccc}
\tablecolumns{10} 
\tabletypesize{\scriptsize}
\tablecaption{Description of UGC 10214 observations}
\tablewidth{0pt}
\tablehead{
\colhead{} & \colhead{} & \colhead{} & \colhead{} & \colhead{} & \colhead{} & \colhead{} &
\multicolumn{3}{c}{Image depth $^b$} \\
\colhead{Filter} & 
\colhead{RA(J2000)} & 
\colhead{DEC(J2000)} &
\colhead{Expt time(s)} & 
\colhead{N exp} & 
\colhead{N orbits} & 
\colhead{Area (arcmin$^2$) $^a$} &
\colhead{Overlapping} &
\colhead{Shallow} &
\colhead{HDF-S} 
}
\startdata
F475W  & 16:06:17.4 & 55:26:46 &  $13600$ & $12$ & $6$ & 11.54 (14.48) &  27.64 (27.86) & 27.27 (27.49) & 27.97 (27.90)\cr
F606W  & 16:06:17.4 & 55:26:46 &  $8040 $ & $12$ & $4$ & 11.54 (14.49) &  27.48 (27.70) & 27.14 (27.36) & 28.47 (28.40)\cr
F814W  & 16:06:17.4 & 55:26:46 &  $8180 $ & $12$ & $4$ & 11.54 (14.46) &  27.03 (27.25) & 26.62 (26.84) & 27.84 (27.77)\cr
\enddata
\label{tab:obs}
\tablecomments{$^a$The values given in this column are the effective (i.e.,
"Tadpole-less") areas used in the analysis, and the accompanying
values in parentheses are the total areas actually covered by the ACS
observations.
$^b$HST image depths taken from \cite{Benitez-etal-03} for point-like
and extended (values in parentheses) sources.}
\end{deluxetable*}

The spheroids in our study were selected from one of the first science
observations acquired by the ACS 
\citep{Ford-etal-02} Wide Field Channel (WFC) as part of the Early
Release Observations (EROs) program. The ACS/WFC houses two $2048
\times 4096$ butted CCDs separated by a $\sim 2''$ gap providing a FOV
of $202''\times 202''$ with a pixel size of $0.05''$.

\subsection{ACS/WFC Observations}

\begin{figure*}
\plotone{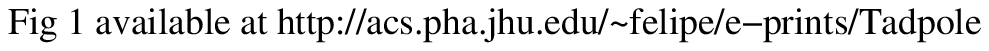}
\caption{Composite color image of UGC 10214 (``The Tadpole''). The
  image illustrates the abundance of background objects observed in
  this deep ACS image. The impressive depth and image resolution is
  shown in the figure insert.}
\label{fig:tadpole}
\end{figure*}

In our analysis we utilize the deep images of UGC10214 (also know as
``The Tadpole'', VV029 and Arp 188), a bright spiral with long tidal
tail at $z=0.03136$, imaged in the F475W$(g)$, F606W$(V)$ and
\mbox{F814W$(I)$} filters. Due to an error in telescope pointing
observation in which the ``head'' of the Tadpole was cut off the ACS
field of view, a second set of observations was performed soon after,
leading to an extra deep exposure on the overlapping region.  In
Table~\ref{tab:obs} we report the observations of UGC 10214, including
exposure times, number of orbits and depth.  The final observations
resulted in a superb multicolor dataset in which the depth limit is
$\simless 1$ mag shallower than the southern HDF (HDF-S) for all the
observed bands.
The properties of the Tadpole galaxy itself and its formation activity
in young stars in the tail have been studied in detail by
\cite{Tran-etal-03}. In this paper, we focus instead on the properties
of a sample of galaxies in the Tadpole field. It is important to point
out that although the apparent large size of the Tadpole galaxy fills
the whole frame, there is a large remaining area of the WFC image of
11.54 arcmin$^2$ which is essentially clean of any foreground
interference and hence perfectly suited for the study of distant field
galaxies in a similar fashion to those performed in the
HDFs. Hereafter we will concentrate on and refer only to the objects
in the background of the WFC image of UGC 10214. In
figure~\ref{fig:tadpole} we show a composite color image of the field
in which it is possible to appreciate the abundance of faint
background galaxies.

\subsection{Image Reduction and Object detection}

\begin{deluxetable}{lcc}
\tabletypesize{\scriptsize}
\tablecaption{ACS Photometric Zero Points in the AB and Vega systems}
\tablewidth{0pt}
\tablehead{
\colhead{Filter} & 
\colhead{AB System} & 
\colhead{Vega} 
}
\startdata
F475W  & 26.043 & 26.144 \cr
F606W  & 26.505 & 26.410 \cr
F814W  & 25.941 & 25.496 \cr
\enddata
\label{tab:zero}
\tablecomments{ACS/WFC photometric zero-points obtained from \cite{Sirianni-etal-04}.}
\end{deluxetable}

The data were first processed by the STScI CALACS pipeline
\citep{Hack-99}, which included bias and dark subtraction, flat
fielding and counts-to-electrons conversion. The images were then
processed by Apsis (ACS pipeline science investigation software) at
Johns Hopkins University. This pipeline which measured rotations and
offsets between dithered images from both pointings, rejected cosmic
rays (CRs) and combined them into single geometrically corrected
images using the drizzle method in each band and a combined image used
for object detection. The reader is referred to
\cite{Blakeslee-etal-03} for a thorough description of Apsis.

Object detection, extraction and integrated photometry were taken from
the SExtractor \citep{Bertin-Arnouts-96} catalogs produced by the
Apsis pipeline. Extensive simulations and tests were conducted to
determine the optimal parameters for extraction and deblending of
ACS/WFC field galaxy sources. A detailed description of the procedure
and parameters is given by \cite{Benitez-etal-03}. In a nutshell,
object extraction was carried out over a specially created detection
image composed of the addition of the images in all bands. Objects
detected above a certain threshold limit within the detection image
are included and selected for subsequent photometry. Apsis provides
--through SExtractor-- a stream of magnitude measurements computed
for each of the bandpasses. For the purpose of our analysis and object
selection we choose the near-total magnitude {MAG\_AUTO} based on
Kron-like elliptical aperture magnitude measured within $k$ times the
isophotal radius. While the SExtractor magnitudes are calibrated in
the AB system as Apsis products, we transformed them into Vega
magnitudes to ease the comparison with previous works, employing the
photometric zero-points from the ACS photometric calibration program
\citep{Sirianni-etal-04}. Hereafter we will refer to magnitudes in the
Vega system. 

\subsection{Selecting Spheroids}
\label{sec:selection}

In selecting field spheroids we first choose objects with integrated
magnitude $I_{814} < 24$, using {MAG\_AUTO} from the SExtractor
products. This is our starting point for segregating early-types. We
decided on selecting objects brighter than $I_{814} < 24$ ---the same
criteria adopted by \cite*{Menanteau-etal-01} in the HDFs--- as this
enables both reliable morphological classification and abundant
signal-to-noise pixel information on deep HST images and will provide
us with the right tools to compare with the HDFs sample later. This
lead us to an initial sample of 373 objects.

\begin{figure}
\plotone{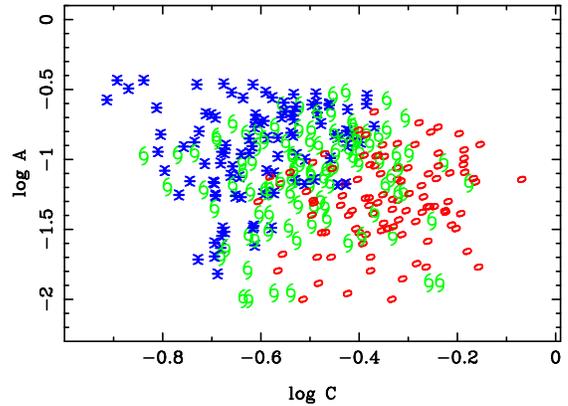}
\caption{The computed values of the Central Concentration and
  Asymmetry and their visual classes. Plot symbols denote E/S0 systems
  as red ellipses, spirals as green spirals and irregulars as blue
  asterisks.}
\label{fig:AC}
\end{figure}

We employ the same procedure for selecting spheroids as described in
\cite*{Menanteau-etal-01} and \cite{Abraham-etal-96} based on a
combination of visual classification and machine morphological
analysis. Stars were initially removed using the SExtractor star class
parameter in addition to visual inspection of all objects in the
sample. Here, we briefly summarize the strategy taken and refer the
reader to these works for a full description of the methodology used.
Galaxies in the Tadpole field were morphologically classified using
an automated classification based on both the central concentration
{\em(C)} and asymmetry {\em(A)} parameters ({\em A -- C}) from
\cite{Abraham-etal-96}, as well as using visual classifications made
by one of us (FM). Visual classification have been shown to agree very
well with {\em A -- C} classes in previous studies
\citep{Brinchmann-etal-98, Menanteau-etal-99}. The segregation of
early-type systems is particularly robust as their chief diagnostic
parameter for discrimination is central concentration. It is
worth noting the key advantage in adding {\em A -- C} to visual
classes alone, as they represent objective measurements of the
morphological properties of galaxies and can be modeled and easily
reproduced in the future if required. In building our final sample of
spheroids we used a combined criteria of using both {\em A - C} and
visual classes, which we believe is a robust approach.
In figure~\ref{fig:AC} we compare the {\em A-C} values computed for
galaxies keyed to their visual classes in three broad categories:
E/S0, spiral and irregular.
We selected spheroid galaxies as objects which were visually
classified as E, E/S0 and S0 in the Hubble scheme. When there were
discrepancies between visual classes and {\em A -- C}, they were
examined in detail, to ensure an unbiased selection. In this fashion
we constructed a final catalog containing 116 ACS spheroids, which is
$\sim43 \%$ larger than the only other HST existing catalog; the HDFs
fields with 79 spheroids.

\subsection{Bayesian photometric redshifts}

Despite the absence of spectroscopic information for galaxies in the
Tadpole field, we enhance the sample with the photometric redshift
estimates computed using the Bayesian Photometric Redshift package
(BPZ) \citep{Benitez-00} included in the Apsis pipeline products. BPZ
follows a Bayesian statistical approach to estimate redshifts
employing a set of galaxy templates and magnitude-redshift
priors. While it is often assumed that several filters are necessary
to achieve accurate measurements, the addition of priors can lead to
reliable redshifts with only the F475W, F606W and F814W filters. As
described in \cite{Benitez-etal-03}, when the Bayesian methodology is
calibrated with the Northern HDF (HDF-N) spectroscopic sample, it has
an excellent behavior at $z<1.5$, without catastrophic outliers and
much better performance than maximum likelihood alone. Furthermore,
according to \cite{Benitez-etal-03}, simulations of the efficiency of
the photometric redshifts as a function of magnitude confirmed that
for bright objects, $I_{814}\simless 24$ the tipical uncertainties are
\mbox{$dz/(1+z)\sim0.1$}, which coincides with the magnitude limit set
for our sample where photometric estimates are most reliable. Despite
the small estimated errors, we opt to take the safer approach of using
the photometric redshift only over the integrated properties of the
sample rather than those of individual galaxies.

\section{Resolving Spheroids}

Understanding the behavior of the point-spread function (PSF) becomes
important when studying distant objects at pixel-scales. In the
present analysis our main concern is that PSF variations as a function
of wavelength may lead to spurious centrally concentrated
inhomogeneities arising from some centrally concentrated profiles in
some spheroids. To study this effect we made extensive simulations of
the effect of the PSF by artificially creating spheroids which were
subsequently analyzed using the same method applied to the observed
data. These are presented in Appendix~\ref{app:psf}. We conclude that
the effect of the PSF will not significantly influence any of the
indicators used in our analysis.

Prompted by the findings of spheroids with central blue cores in the
HDFs \citep*{Menanteau-etal-01}, the first step in our analysis is
to construct $V_{606}-I_{814}$ color maps for all the ellipticals in
our sample. This color peculiarity has only been reported for
early-types in the HDFs so far, and it is compelling to verify whether
it will also be present in other unrelated samples such us the current
one. For this, we compute pixel maps, surface brightness, $\mu$, for
all galaxies in all three bands, using the well-known transformation,
\[ \mu = {\rm zp} - 2.5\log({\rm counts}/{\rm exptime}) + 5\log({\rm pixelscale}) \]
where zp is the zero-point for a given filter in the Vega system from
Table~\ref{tab:zero}, the pixelscale is 0.05 arc/pixel and counts is
the number of electrons in each pixel. Subsequently, color maps were
computed, selecting only contiguous pixels above a surface brightness
threshold of \mbox{$\mu_{606}\leq25$~mag/arc sec$^2$}. We note that we
use this surface brightness limit only in the computation of  \dvi\
in section~\ref{sec:dvi} in order to maintain consistency when
comparing with previous observations. However in
section~\ref{sec:slope} we will discusss a more robust method for
selecting and measuring the light of spheroids.

\subsection{Old ellipticals vs Blue ellipticals}

\begin{figure}[t!]
\plotone{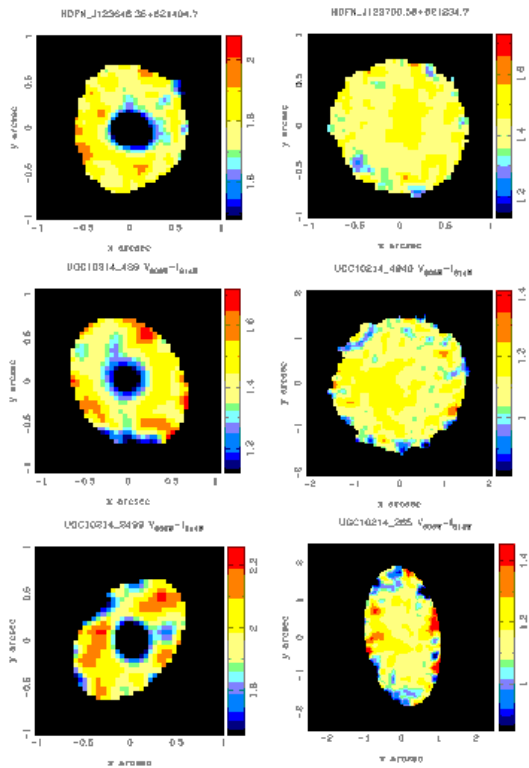}
\caption{Comparison of the \vi internal color properties of spheroids
  in the Tadpole field and in the HDF-N. The top two galaxies are taken
  from the HDF-N sample and the ones below, from the ACS sample. The
  left column shows examples of blue nucleated spheroids and on the right
  normal spheroids.}
\label{fig:comparison}
\end{figure}

Our first objective is finding a population of spheroids with blue
cores and inhomogeneous internal colors in the current ACS sample
mirroring those found in the HDFs. After an initial inspection, we
conclude that in most cases spheroids show fairly smooth and
homogeneous \vi color maps, but the existence of a subsample with
perturbed internal colors is quite evident. We report a fraction of
spheroids with blue cores comparable to the HDFs in our ACS sample,
with similar sizes and strength in their color differences. It is
important to remark that this is the first sample, apart from the HDFs
in which resolved studies of distant galaxies are possible. To
illustrate the similarities in our findings, in
Fig.~\ref{fig:comparison}, we show examples of spheroids with normal
and peculiar internal color properties in the HDF-N and the present
sample of ACS spheroids. The similarities in the form of the blue
nucleated ones in both samples are quite evident.

\section{Methodology and analysis}

After considering the color maps of spheroids, we want to investigate
how the internal colors of field ellipticals vary as a function of
redshift and quantify the fraction that departs from a passive
evolving population. Although we can identify a distinctive population
of spheroids with blue cores and color inhomogeneities from visual
inspections alone, we wish to develop a method for quantifying its
strength in an objective way.
A standard tool for this purpose is to model their color properties
using stellar synthesis population models
\citep{Abraham-etal-99,Menanteau-Jimenez-01}. Unfortunately, these
predictions are quite sensitive to the galaxy redshift of observation
and our present sample only encompass photometric estimates, making
this approach toward characterizing individual galaxies somewhat
uncertain. Instead we chose to explore a model-free approach and focus
on the integrated properties, which in our view represent a safer way
to evaluate the trend of the population.
For this purpose we revisit the internal color dispersion estimator
from \cite*{Menanteau-etal-01} and devise a new measurement based on
the slope of the color gradients of a galaxy that we will show is more
sensitive to the presence of central blue spots.

\subsection{The internal color dispersion \dvi}
\label{sec:dvi}

The idea of studying the internal colors of an individual galaxy can
be considered as a generalization of the use of the color-magnitude
relation in the study of the history of elliptical galaxies in
clusters \citep*{BLE92}, where the photometric dispersion is employed
as a powerful mechanism to determine variations in their star
formation history.
The first quantity that we present is the signal-to-noise weighted
scatter in the internal \vi colors of a galaxy, \dvi, as defined in
\cite*{Menanteau-etal-01}. This measurement has proven a good tracer
of peculiarities in spheroid colors at pixel scales in the sense that
objects with low values of \dvi\ have very uniform and normal color
maps, while those with higher values do present disturbed internal
colors and in many cases blue cores.

A key motivation for re-visiting this estimator is that it has already
been applied and calibrated with a similarly selected sample of
galaxies in the HDFs, where it proved to be a good discriminant of
objects with blue cores vs ellipticals with normal colors. This
feature enable us to make direct comparison between the HDFs sample
\dvi\ measurements and the current one. In addition, it is a model
independent method for studying differences in the star-formation
histories of galaxies, making it free of assumptions regarding the
epoch of formation in ellipticals.

In our estimates of \dvi, we follow the same procedure used in the
HDFs, and we also chose to concentrate only on the \vb and \ib bands
for computing color dispersions, as the resulting \vi color contains
significantly smaller observational errors than \bv, specially for
systems dominated by old stellar populations \citep*[see][for
details]{Menanteau-etal-01}. However, it is important to note that
strong variations in the internal colors are also present when looking
at the the \bv colors.
We select all contiguous pixels within a
surface brightness limit of \mbox{$<25$~mag/arcsec$^2$} in $V_{606}$, to
isolate the galaxy from the background sky and maximize the S/N ratio
per pixel associated. Using only the pixels within this limit, \dvi\ is
computed from the weighted signal-to-noise distribution of \vi colors
for individual pixels, which in turn identifies the internal
homogeneity of a galaxy, using the following recipe:
\begin{equation}
\delta(V-I) \propto \frac{\sum ( x_{i} -
  \bar{x})^{2}S(x_i){\rm{SNR}}(x_i)}{\sum S(x_i){ \rm{SNR}}(x_i)}
\end{equation}
\begin{equation}
\bar{x} = \frac{\sum x_{i}S(x_i){\rm{SNR}}(x_i)}{\sum S(x_i){\rm{SNR}}(x_i)}
\label{dvi-eqn}
\end{equation}
where $S(x)$ is a selection function for pixels with signal/noise
above a certain threshold such that $S(x)=1$ for pixels above and
$S(x)=0$ for pixel below the threshold. $SNR(x_i)$ is the
signal-to-noise ratio for a given pixel color $x_i$. The selection
function $S(x)$ and the weighting according to $SNR(x)$ address biases
arising from noise variations at pixel-scales by rejecting low signal
pixels and weighting the pixels' contribution proportionally to their
signal.

\subsection{The mean \vir\ slope of the galaxy}
\label{sec:slope}

In addition to the color dispersion we wish to develop an alternative
quantity to isolate spheroids with non-homogeneous internal
colors. This should be suited to measure the strength of the blue
cores and not prone to dispersions arising from particularly reddened
core.  For this we concentrate on a method that quantifies the
variation in the color as a function of the galaxy radius. Given that
in most of the cases, color peculiarities manifest themselves via the
presence of blue cores, a clear pattern of this effect is the display
of inverted (i.e. positive sign) color gradient, as opposed to
``normal'' passively evolving ellipticals which either show flat or
slightly negative color gradients, as expected from metal enriched
cores.

We focus on the \vi colors to compute color gradients as a function of
the radius $r$, as these represent the bands where the signal-to-noise
is higher and can be more directly related to \dvi.  In the extraction
of the color gradients, initially we measure the centroid and second
order moments ($\overline{x^2}$, $\overline{y^2}$, $\overline{xy}$) of
the galaxy utilizing the $I_{814}$ band, from which we determine the
ellipticity parameters of the objects ($a, b, \theta$).  It is worth
noting that we keep the same elongation ratio $e = a/b$ up to the
isophotal limit to which the color gradient is computed. Next, using
concentric ellipses, we calculate the $V_{606}-I_{814}(r_i)$ gradients
as the median color in the shell between $r_{i-1}$ and $r_{i+1}$. To
gauge the slope of the color gradient, we fit a fourth order
polynomial function to the observed \vir, up to a certain maximum
radius $r_{max}$, from where we measure the slope by simply obtaining
the first derivative of the fitted polynomial function. 
In order to consistenly measure the slope of galaxies over similar
physical areas on spheroids and at different redshifts, we compute the
slope up to a radius $r_{max}$ following a similar aproach to the one
described by \cite{Papovich-etal-2003} and
\cite{Conselice-etal-2000}. This consists of choosing the aperture
radius $r_{max}$ of the galaxy as $r_{max} = 1.5\times r_p$ with
$(\eta=0.2)$ where $r_p$ is the \cite{Petrosian-76} radius and
$\eta(r)$ is defined as:
\begin{equation}
\eta(r) = I(r) / \langle I(r)\rangle 
\end{equation}
where $I(r)$ is the surface brightness of the galaxy in an annulus of
radius $r$ and $\langle I(r)\rangle$ is its mean value up to the same
radius. The Petrosian radius depends only on the galaxy surface
brightness and it is independent of the redshift of observation.
Formally, the mean \vir\ slope
of the galaxy can be expressed as:
\begin{equation}
\overline{slope} = 
\frac{\sum  
   \frac{d}{dr}(V-I)(r) |_{r_i} \cdot \Delta r_i } {\sum \Delta r_i}
\end{equation}
where the derivative is taken over the fitted function, $\Delta r_i$
is the size of the concentric shells and $r$ is measured in kpc. We
repeat this procedure to all the galaxies in our sample. In
Section~\ref{sec:results} we will investigate how it performs compared
with \dvi\ and present examples.

\subsection{Comparing the indicators}

An important objective of our analysis, is to carry out direct
comparisons of the internal color dispersions between the present ACS
sample and those observed in the HDFs. For this purpose, we will
include in our study the morphologically selected catalog of spheroids
in HDF-N from \cite{Menanteau-etal-01}. The HDF-N sample was
constructed using the same selection limits and morphological
classification, ensuring a robust comparison between samples. We
decide to include only the Northern HDF as this contains the largest
number of spectroscopic redshifts, making possible an investigation of
how the internal color dispersions vary as a function of redshift as
opposed to the photometric estimates in ACS spheroids. After computing
\dvi\ and the mean slope for all galaxies in the ACS and HDF-N samples
we examine how they perform against each other and subsequently
compare with HDF-N spheroids.

\begin{figure}
\plotone{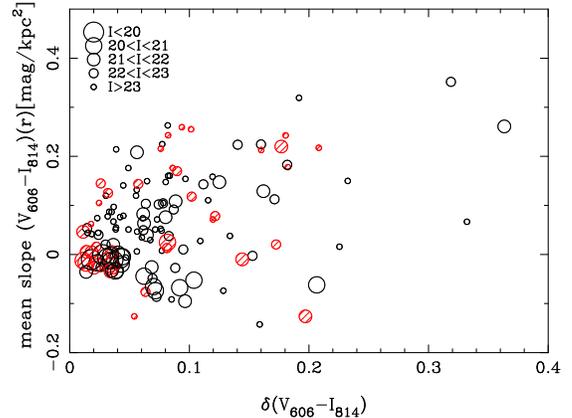}
\caption{Comparison between the two statistics used to probe
  variations in the internal color properties, \dvi\ and the mean
  slope $(V-I)$ for spheroids in the Tadpole field, represented by
  open circles. Values for the HDF-N are also shown as hatched
  circles. The size of the symbols is keyed to the \ib magnitude as
  indicated in the panel within. Galaxies with large positive mean
  slope \vir\ bear blue cores, while those with large \dvi\ values
  show strong color dispersion in general associated with blue cores.}
\label{fig:slope-dVI}
\end{figure}

Figure~\ref{fig:slope-dVI} shows the values of \dvi\ versus the mean
slope \vir\ for all spheroids in the Tadpole field. We can observe a
clear correlation between both statistics for the larger part of the
sample: galaxies with high positive mean slopes (indicative of blue
cores) also show high values of \dvi\ (i.e. high internal color
dispersions). Although the inverse relation is true for most galaxies
(i.e high color dispersion values (\dvi$\simgreat0.06$) also have high
positive slopes) we see in Fig.~\ref{fig:slope-dVI} that several high
\dvi\ galaxies have flat or even negative slopes in some cases. This
results as galaxies with metal enriched cores tend to have redder
centers, making their mean slopes slightly negative or flat, while
keeping higher than normal values of \dvi.
From Fig.~\ref{fig:slope-dVI}, we wish to highlight that the mean
slope \vir\ is a more robust statistic for tracing blue cores as
spheroids with positive slopes and hence blue cores exhibit high
internal color dispersion.
Also in Fig.~\ref{fig:slope-dVI} we present the values computed for
the HDF-N spheroids. It is reassuring that ellipticals populate the
same region in the \dvi---median slope space for both samples. After
individually inspecting the objects with high \dvi\ values but low or
negative slopes, we confirm that these correspond to objects with
reddened cores. We conclude that although both statistics succeed in
tracing the observed color variations, the mean slope is a more robust
statistic for tracing blue light in spheroids and is not prone to
color differences arising from color gradients normally accounted for
in passively evolved ellipticals.

\section{Results}
\label{sec:results}

Establishing the existence of a population of field spheroids with
marked color dispersions, resembling those exhibited by HDFs spheroids
is a major objective of this paper. From the constructed color maps
alone we can conclude that this is a property shared by field
spheroids at intermediate redshifts regardless of the sample in
study. The presence of strong variations in the internal color
distributions suggests episodes of recent start-formation
activity. However, detailed measurements regarding the intensity and
timescales involved requires precise spectroscopic information, not
available for the sample in our study. Instead, we opt to determine in
a model-independent way the proportion of active systems as a
function of redshift and based on this information, make educated
guesses regarding the evolution of early-type galaxies.

\begin{figure}
\plotone{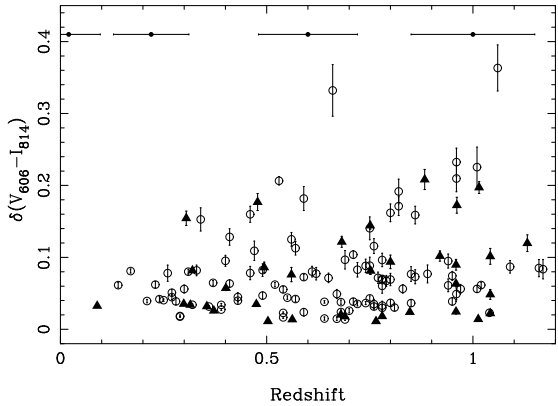}
\caption{The scatter \dvi\ as a function of redshift for spheroids in
the Tadpole field are shown as open circles. For reference the values
for spectroscopic redshifts from the HDF-N are also shown as filled
triangles. Horizontal bars at the top represent typical 1$\sigma$
photometric redshift errors for galaxies in the ACS sample.}
\label{fig:redshift-dVI}
\end{figure}

\begin{figure*}
\plotone{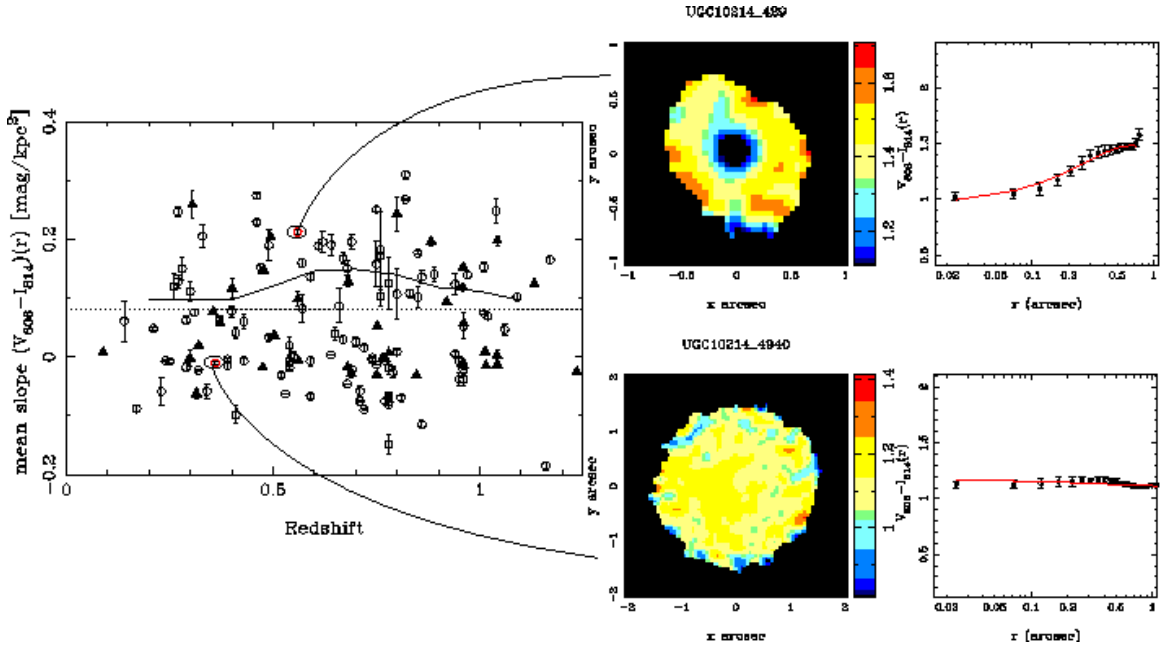}
\caption{The mean slope \vir\ as a function of redshift for spheroids
in the Tadpole field are shown as open circles. For reference the
values with spectroscopic redshifts from the HDF-N are also shown as
filled triangles. The dotted line is the limit used for selecting
evolving systems as a function of redshift. To illustrate how the mean
slope \vir\ traces core colors, we show two examples of objects with
positive/high (top) and negative slopes (bottom), with their
associated position in the mean slope vs redshift space. We also
display their \vi color maps and their corresponding \vir\ extracted
gradients in the far right panels, with their associated $1\sigma$
error bars. The solid red line in the extracted color gradient
represents the fitted slope used in the computation of the mean
slope. The solid line shows the mean slope values obtained for
simulated galaxies as a function of redshift.}
\label{fig:redshift-slope}
\end{figure*}

Initially, we wish to investigate the behavior of \dvi\ as a function
of redshift. For this, in figure~\ref{fig:redshift-dVI} we display the
computed values of \dvi\ as a function of their photometric
estimations for the Tadpole field (open circles) and spectroscopic
redshifts for the HDF-N (filled triangles). Error bars represent
$1\sigma$ estimates obtained using the bootstrap resampling technique
introduced by \cite{Efron-Tibshirani-86}. From
figure~\ref{fig:redshift-dVI} we observe how the fraction of spheroids
with high \dvi values slightly increases as a function of redshift in
the same way as reported for HDF-N spheroids. We make a quantitative
measurement and focus on the fraction of systems with high \dvi values
compared to low ones as this measurement is not prone to selection
effects due to the redshift distribution of spheroids in our
sample. We separate the sample into two redshift bins, such as
$0<z_{bin1}\leq0.6$ and $0.6< z_{bin2} \leq1.2$ and calculate the
fraction of spheroids with $\delta(V-I) > 0.09$ to the total
number. We report an increase from 28\% in $z_{bin1}$ to 39\% in
$z_{bin2}$. However, we point out that this might also be consistent
with no increase given the relatively small number of objects
involved.

It is reassuring to confirm that both samples show a very similar trend
in the observed spread as well as its increase with redshift. Assuming
that in most cases high values of \dvi\ are associated with recent
star formation activity within the galaxy, the rise of \dvi\ with
redshift indicates a clear trend in this sense. For the HDFs,
\cite{Menanteau-etal-01} have computed simple models that can explain
the existence of galaxies with high \dvi\ as the result of small
additional bursts of star formation activity.

If we assume that strong internal color dispersion are indeed
associated with recent star formation, it is interesting to obtain the
fraction of evolving systems as a function of redshift. As observed
from figure~\ref{fig:slope-dVI}, the mean slope \vir\ is better suited
to probe blue cores and internal inhomogeneities than $\delta(V-I)$,
therefore we choose to use this instead of \dvi\ for estimating the
ratio of spheroids with strong internal color variations to the total
number of objects. In figure~\ref{fig:redshift-slope} we show computed
values of the mean slope \vir\ as a function of redshift for both the
Tadpole field and the HDF-N sample, using the same symbols as in
figure~\ref{fig:redshift-dVI}. For completeness we also display the
values for the HDF-N, as this allows to make consistency checks for
variations that could arise between samples, in part due to the use of
photometric redshifts. Error bars show $1\sigma$ computations
calculated by carrying out Monte Carlo simulations over the data. To
illustrate how the mean slope is used in probing galaxies with blue
cores, in figure~\ref{fig:redshift-slope} we also show two galaxies of
contrasting internal colors and the very distinctive values of the mean
slope they have. It is interesting to note from this figure how
galaxies tend to populate rather uniformly the redshift space.

\subsection{The mean slope \vir\ redshift dependence}

In order to reliably make use of the mean slope \vir\ in tracing blue
light in spheroids, we need to investigate its variation as a function
of redshift introduced from measuring the mean slope using the \vb and
\ib filters. We attempt to reproduce through simulations the observed
spatially resolved color properties of spheroids with blue nuclei. Our
key assumption when modeling these spheroids is that at intermediate
redshifts they can be described using a two component model: a young
stellar central component (responsible for the blue light)
superimposed on an old system. We modeled the integrated \vi colors of
both stellar component using spectral energy distributions (SED) from
the Bruzual \& Charlot (1996) (BC96) library. For the old stellar
component we chose an SED corresponding to exponentially declining star
formation ($e$-folding time $1$~Gyr) at age $12$~Gyr, which resembles
the colors of a normal elliptical at $z\sim0$. For the central region
we employ the SED of a recent star-forming elliptical as observed
after $1$~Gyr since its formation. For each component their observed
\vi colors and apparent magnitudes are computed as as a function of
redshift ($K$-correction only) assuming a flat universe
($H_0=71$~km~s$^{-1}$Mpc$^{-1}$, $\Omega_0=0.27$,
$\Omega_\Lambda=0.73$).

To emulate the geometrical properties of the modeled galaxies (as
those of the blue nucleated spheroid in
Fig.~\ref{fig:redshift-slope}) we created artificial ellipticals from
the two components using a customized version of the {\sc IRAF}
package {\sc ARTDATA}. We used a fixed physical size of half-light
radius $r_e = 3$~kpc for a de Vaucouleurs profile for the old
component. We assume that the blue central component corresponds to
only $15\%$ of the total stellar mass of the galaxy
$M=3\times10^{11}M_{\sun}$, as this supplies enough blue light to
reproduce the observed blue colors, and effective radius of $0.15r_e$.

For the resulting artificial galaxies we compute the mean slope \vir\
between $0.2<z<1.2$ using the same procedure as for the real data. We
show the results of the exercise in
figure~\ref{fig:redshift-slope}. The solid line represents the
computed values of the mean slope \vir\ recovered from the
simulations. We observe that for simulated spheroids the mean slope
changes little as a function of redshift, with a bump at $z\sim 0.65$
which coincides with the peak in the $\Delta(V_{606}-I_{814})$ colors
difference between both components. We conclude that although our
filter set is more favorable to trace blue light near $z\sim0.65$, the
variations as a function of redshift are small enough that this will
not significantly affect our results.

\subsection{The fraction of active systems}

\begin{figure}
\plotone{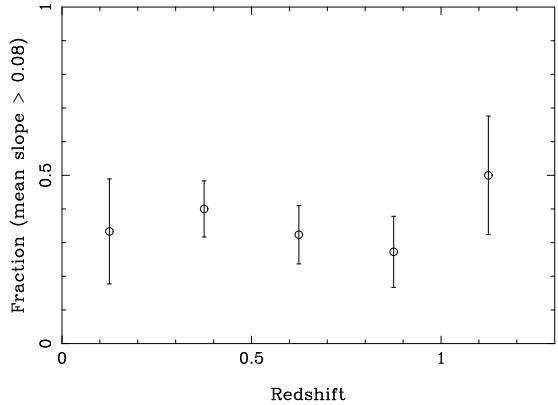}
\caption{The fraction of active systems to total as a function of
  redshift for spheroids in the Tadpole field. Error bars represent
  $1\sigma$ estimates obtained using Monte Carlo simulations. The
  value for the first redshift bin is inprecise, as the sample is
  incomplete for low redshift galaxies.}
\label{fig:fraction}
\end{figure}

When computing the active to total galaxies ratio, we need to
calibrate the mean slope \vir\ to differentiate between active versus
passively evolved systems. Certainly high values of the mean slope are
directly linked to large internal color variations, and for the high
mean slope vs. redshift envelope, is where undoubtedly the most
striking cases of blue cores arise. On the other hand, galaxies with
small values, near zero or negative show no significant color
variations. However, defining a limit that separate active vs passive
systems is a rather uncertain and imprecise exercise. To assess this
problem, we opt for interactively calibrating this limit, by visually
inspecting all galaxies and their respective color maps. After
extensive inspections, we resolve to chose a constant limit of mean
slope \vir\ $>0.08$ mags$/$kpc$^2$ as a fiducial limit, that produces
the cleanest separation (Fig.~\ref{fig:redshift-slope}. Using this
limit we proceed and compute the fraction of active to total spheroids
as a function of redshift for all the ellipticals in the Tadpole field
up to $z\sim 1.2$. In order to avoid possible uncertainties arising
from normal Poisson variations and the use of photometric redshifts, we
opt for binning out the data using redshift bins of size $\Delta
z=0.25$, wide enough to smooth out major noise variations. In
figure~\ref{fig:fraction} we show the fraction of active to total
systems as a function of their photometric redshifts. In order to
incorporate the error in the measurement of the photometric redshifts,
$1\sigma$ error bars in figure~\ref{fig:fraction} were estimated by
carrying out Monte Carlo simulations over the distributions. To
estimate the uncertainties we recompute the histogram $N$ times
($N\sim1000$) in which the redshift of observation of the galaxies is
modified such as $z_{o}' = z_o \pm \alpha\delta z$, in which $\alpha$
is a randomly generated number between $[-1,+1]$ and $\delta z$ is the
nominal error estimation of BPZ equal to $\delta z=0.15(1+z)$.

From figure~\ref{fig:fraction} we report that the fraction of active
spheroids is compatible with constant as a function of redshift
(within the error bars). It is important to notice that the estimate
for the lowest redshift bin, is quite uncertain as indicated by the
large error bars, as the Tadpole sample is incomplete for lower
redshift galaxies \citep[see][]{Benitez-etal-03}. Regarding the actual
amount of the fraction of active spheroidals, we recognize that its
absolute value is rather dependent in the limit used for
selection. However, for the current limit adopted, this number mostly
fluctuates around $30\%-40\%$ in good agreement with the values
obtained for the HDF-N for \cite{KBB99} and
\cite{Menanteau-etal-01}. It is also worth noting that from tests
performed on the sample, the shape of fraction distribution remains
largely unchanged when adopting a different limit.

It is appealing to analyze our estimate of the number of active
spheroids in the context of their evolutionary history using the
simple methodology introduced in this paper. The relatively constant
to small rise in the fraction of active systems is suggestive of the
presence of a population of active spheroids, continuous at least up
to $z\sim1$. This can also lead to the interpretation that spheroids
are being constantly assembled since $z\sim1$. At the same time, the
presence of spheroids with homogeneous colors and smooth color
gradients is clear evidence of the presence of a relatively dominant
population of normal evolved spheroids at all redshifts.

It is possible to ask then, what type of objects are the blue spheroids
that we detect? And, what would be their low redshift counterparts? A
possible explanation, might be that we are seeing these galaxies as
they enter into the E/S0 class, possibly within the first Gyr, since
their last star-formation activity. From the modeling of stellar
synthesis populations \citep{Menanteau-etal-01, Menanteau-Jimenez-01},
it is believed that the existence of such populations is not
long-lasting. We might be seeing the predecessors or proto-ellipticals,
in the last period of star-formation before becoming normal
ellipticals as we know them at low redshifts and in clusters of
galaxies.

\section{Conclusions}

In this paper, we have used newly available data from the Advanced
Camera for Surveys to study the internal color properties of sample of
116 morphologically selected spheroids. This represents the first deep
dataset of spheroids available from the ACS, and we used it to
investigate its superb ability to resolve distant galaxies.

Using an independent sample of ACS spheroids we have confirmed the
presence of a population of spheroids with color inhomogeneities
similar to those found in the HDFs. Using a model independent
approach, we have introduced a new statistic, the mean slope \vir\
which probes successfully the presence of blue cores and internal
color inhomogeneities in spheroids. We compare our measurements with
the HDF North with consistent results.
Assuming that strong variations in the internal color dispersion of
spheroids are linked to recent episodes of star formation, we use the
mean slope to separate active and passively evolved systems as a
function of redshift, and based on this, we estimate the fraction of
active systems versus redshift. We found that within the uncertainties
of our measurements, the fraction of active galaxies can be described
as constant, with a trend to grow with redshift, with nominal values
between $\sim30\%-40\%$ of the sample, consistent with previous
results. We take this as evidence for the continuous formation of
spheroids since $z\sim1$, while the data also shows the presence of a
population of old passively evolved ellipticals at all redshifts.

An important part of the challenge that lies ahead is try to
understand the significance of the blue spheroids and whether or not
these represent the early stages of what we expect to be an elliptical
as defined at $z\sim0$. Additionally, alternative theories of galaxy
formation need to be explored which can explain at the same time the
presence of passive evolving systems and recent star-forming
spheroids. It also remains to be learned what are the detailed internal
properties of ellipticals in clusters are from upcoming ACS observations
at $z\sim1$ (Postman et al. 2004).

\acknowledgments

ACS was developed under NASA contract NAS 5-32864, and this research
is supported by NASA grant NAG5-7697. We would like to thank the
anonymous referee for his/her comments and suggestions which improved
the paper.

\bibliographystyle{apj} 

\appendix
\section{The effect of the ACS/WFC point-spread function}
\label{app:psf}

\begin{figure}[t!]
\plotone{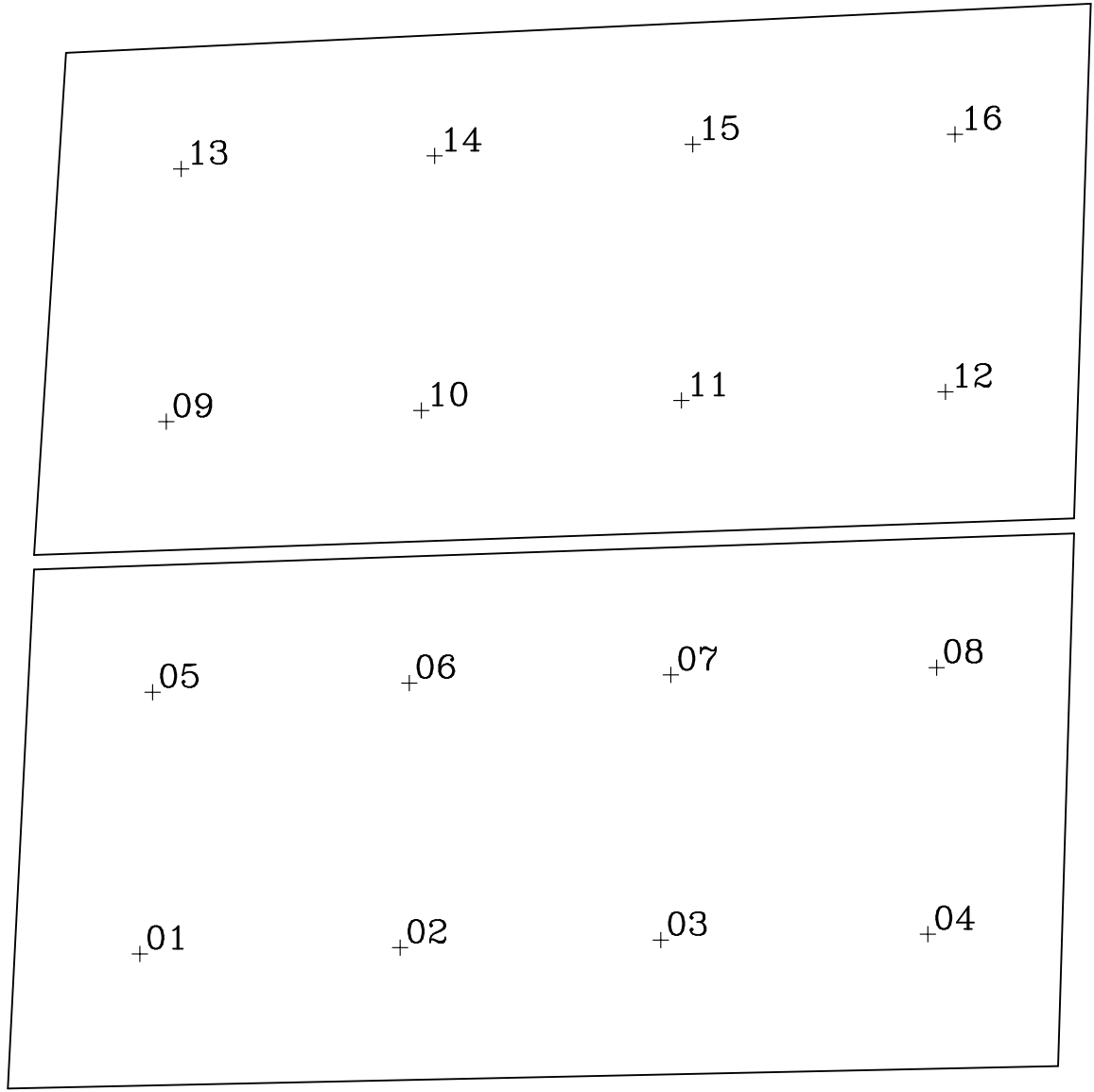}
\caption{The distribution of synthetic PSFs created using Tiny Tim as
  function of the position on the geometrically corrected mosaiced
  ACS/WFC field. The plus symbols show the position of the synthetic
  PSFs in the WFC field.}
\label{fig:shape}
\end{figure}

\begin{figure}
\plotone{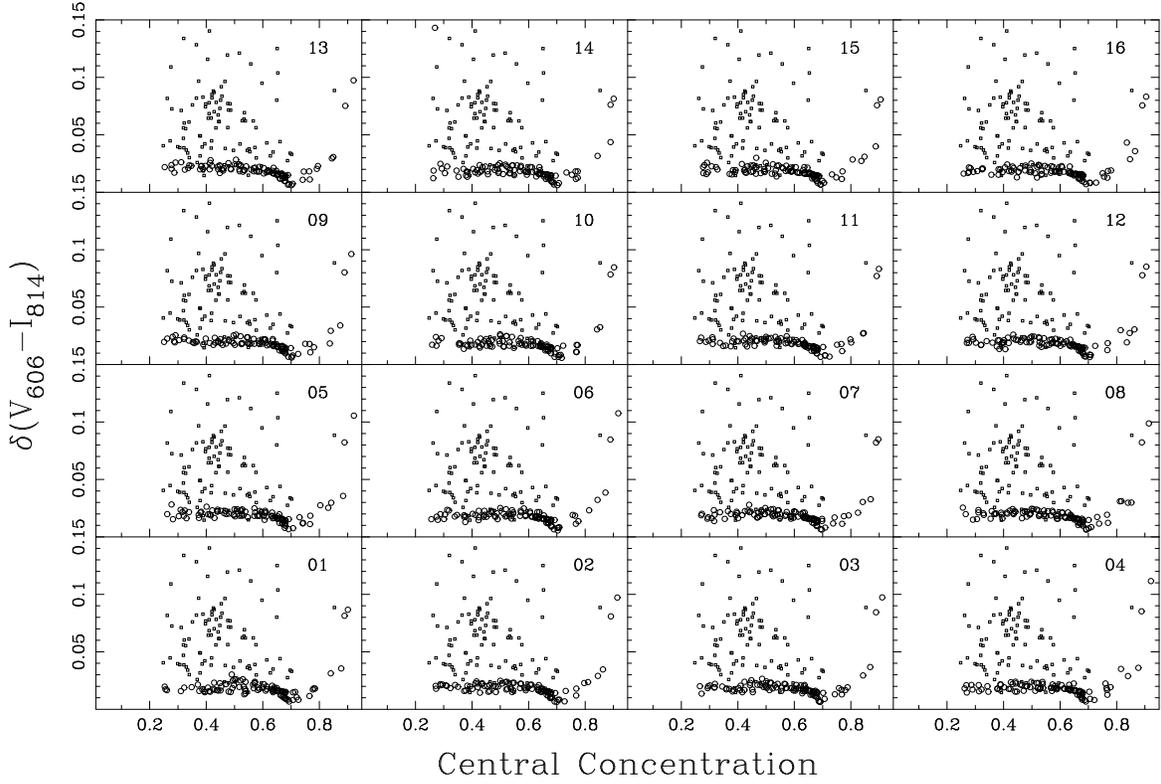}
\caption{The internal color dispersion \dvi, used to measure color
  inhomogeneities, as a function of the central
  concentration, $C$. Open circles represent simulated spheroids to
  estimate the bias level introduced by the PSF.  Open squares
  observed ones. Each panel shows the results of artificial spheroids
  for each PSF positions using the description of
  Fig.~\ref{fig:shape}.}
\label{fig:psf-dVI}
\end{figure}

\begin{figure}[h!]
\plotone{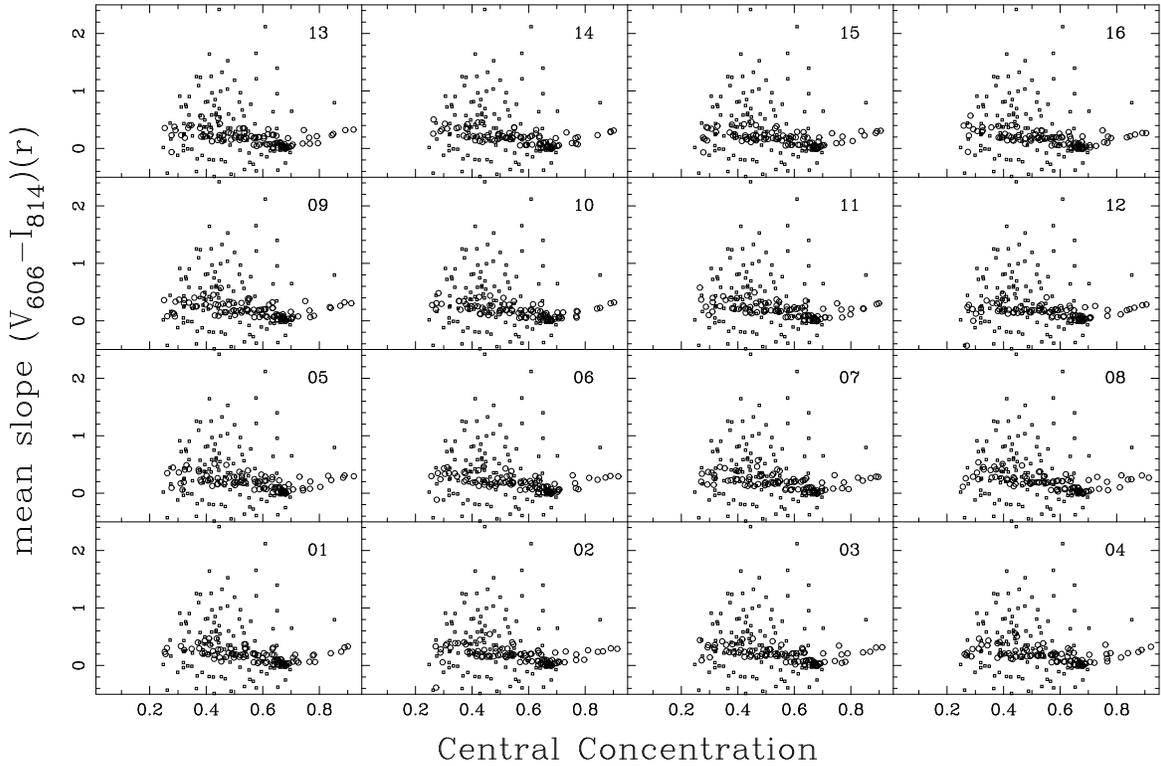}
\caption{The mean slope \vir, used to trace central blue light, shown
  as a function of the central concentration, $C$. Open circles
  represent simulated spheroids to estimate the bias level introduced
  by the PSF.  Open squares observed ones. Each panel shows the
  results of artificial spheroids for each PSF positions using the
  description of Fig.~\ref{fig:shape}.}
\label{fig:psf-slope}
\end{figure}

It is known that systematic variations present in the HST PSF may
influence studies which depend on small aperture photometry or small
galaxies with bright cores. Our main concern is the PSF variation as a
function of wavelength, which might induce spurious centrally
concentrated color inhomogeneities on sharply peaked light
profiles. To investigate the bias level introduced by variations of
the PSF in our internal color analysis of spheroids, we carried out
Monte Carlo simulations following a procedure similar to the one
described in \cite{Menanteau-etal-01}. This consists of creating
artificial galaxies using the IRAF package ARTDATA resembling the
sizes and brightness as of those in our study. These were subsequently
analyzed in the same fashion as the observed data. Initially we
attempted to create artificial de Vaucouleurs profiles using a
combination of averaged observed bright stars in the background of the
Tadpole field as PSF. However, it was soon evident that there was a
systematic variation in the PSF FWHM as a function of position in the
WFC field. In addition, the small number of bright stars in The
Tadpole field made the use of stars as real PSFs too challenging. The
PSF variation across the WFC field has been extensively documented by
\cite{Krist-03}. Although the PSF variations in ACS are less than
other HST cameras, its does varies over the field, most notably for
the WFC (see Fig.~1 in \citealt{Krist-03}). Therefore, we chose to
utilize the synthetic PSFs generated with the Tiny Tim software as it
does take into account field variations.

We created a grid of $4\times4$ PSFs at evenly distributed positions
across both WFC chips with Tiny Tim for the \vb and \ib bands. These
were incorporated into ACS mock raw images and later processed through
Apsis using the same parameters as for the observed data. These yield
a geometrically corrected grid image of synthetic PSFs. In
Fig.~\ref{fig:shape} we show a diagram with the final positions. Using
this grid, spheroids were synthesized at each PSF location with de
Vaucouleurs profiles using IRAF's ARTDATA. A set of simulated
spheroids were generated for the \vb and \ib bands trying to mimic as
much as possible the observed properties of spheroids. These include a
range of half-light radius ($r_e$) from $0.1''-0.7''$ which correspond
to a physical length of $\sim0.75-5.5$~kpc
($H_0=71$~km~s$^{-1}$Mpc$^{-1}$, $\Omega_0=0.27$,
$\Omega_\Lambda=0.73$) and a range of magnitudes consisted with our
sample ($18<I_{814}<24$). We processed the synthetic spheroids using
the exact same procedure employed for real spheroids and calculated
both estimators, \dvi\ and the mean slope \vir\ for the whole
sample. To simplify the simulation we compute the mean slope as
function of the object apparent size. As the influence of the PSF in
the resolved colors is expected to be a strong function of the
'peakiness' of light distribution of the galaxy, we probe the
influence of the PSF on real and simulated galaxies against the
central concentration parameter (section~\ref{sec:selection}).

The results of this exercise are shown in Figs.~\ref{fig:psf-dVI}
and \ref{fig:psf-slope}, where we plot the values of \dvi\ and the
mean slope \vir\ as a function of the central concentration
respectively for the 16 PSF pointings. In all 16 positions we also
plot the observed data (open squared) to compare with the simulation
results. When confronting real and simulated values of \dvi\ in
Fig.~\ref{fig:psf-dVI} we can see that over the observed range of
$C$ for the ACS spheroids (i.e. $0.3\simless C \simless 0.7$) the
values of \dvi\ for the simulated galaxies as significantly lower that
those obtained for the observed spheroids. Only for values of $C>0.8$
we enter into the regime in which the PSF can affect the recovered
values of \dvi\ as the PSF becomes important. However, our spheroids
do not lie in this range. We notice some variation depending on the PSF
position used, but this effect is too small for the scale of the real
values of \dvi. We also compare the central concentration with the
mean slope \vir\ using the same symbols as of
Fig.~\ref{fig:psf-dVI}. The simulated galaxies have mostly a uniform
distribution of values of mean slope \vir\ as a function of $C$, also
lower than the observed for the real data. We conclude that the bias
level introduce by the variations in the WFC PSF will not seriously
affect any of our conclusions.

\section{The Catalog of Spheroids in UGC102104}

\begin{deluxetable}{rrrccrccrc}
\tablecolumns{10} 
\tabletypesize{\scriptsize}
\tablecaption{The catalog of selected spheroids from UGC102104}
\tablenote{Complete table available at {\tt http://acs.pha.jhu.edu/$\sim$felipe/e-prints}}
\tablewidth{0pt}
\tablehead{
\colhead{ID} & 
\colhead{RA(J2000)} & 
\colhead{DEC(J2000)} &
\colhead{F814W} & 
\colhead{\dvi} & 
\colhead{mean slope \vir} & 
\colhead{BPZ} &
\colhead{A} &
\colhead{C}
}
\startdata
 1041 & 16:06:16.49 & +55:23:44.19 & 21.877 & $0.0621 \pm 0.0019$ & $-0.0587 \pm  0.0012$ &  0.23 &  0.054 &  0.538 & \\
 1076 & 16:06:03.55 & +55:24:54.00 & 22.982 & $0.0146 \pm 0.0072$ & $ 0.0292 \pm  0.0335$ &  0.67 &  0.032 &  0.392 & \\
 1103 & 16:06:09.86 & +55:24:28.60 & 21.767 & $0.0393 \pm 0.0026$ & $ 0.0485 \pm  0.0026$ &  0.21 &  0.043 &  0.385 & \\
 1109 & 16:06:19.18 & +55:24:17.63 & 23.713 & $0.1091 \pm 0.0285$ & $ 0.1515 \pm  0.0350$ &  0.47 &  0.067 &  0.275 & \\
 1142 & 16:06:06.01 & +55:24:53.35 & 23.408 & $0.1284 \pm 0.0277$ & $-0.0982 \pm  0.0386$ &  0.41 &  0.047 &  0.365 & \\
 1157 & 16:06:13.95 & +55:24:57.49 & 21.760 & $0.0964 \pm 0.0086$ & $-0.0765 \pm  0.0030$ &  0.78 &  0.052 &  0.466 & \\
 1200 & 16:06:04.61 & +55:25:58.07 & 23.408 & $0.5834 \pm 0.0079$ & $-0.2479 \pm  0.0304$ &  0.85 &  0.051 &  0.477 & \\
 1207 & 16:06:01.80 & +55:26:17.99 & 20.447 & $0.1038 \pm 0.0013$ & $-0.0751 \pm  0.0009$ &  0.71 &  0.081 &  0.653 & \\
 1266 & 16:06:18.02 & +55:24:31.97 & 23.362 & $0.0802 \pm 0.0219$ & $ 0.1880 \pm  0.0419$ &  0.61 &  0.004 &  0.451 & \\
 1275 & 16:06:19.02 & +55:24:25.81 & 22.822 & $0.0882 \pm 0.0080$ & $-0.0050 \pm  0.0077$ &  0.74 &  0.041 &  0.425 & \\
 1282 & 16:06:15.43 & +55:24:49.85 & 23.337 & $0.0563 \pm 0.0151$ & $ 0.1083 \pm  0.0159$ &  0.83 &  0.130 &  0.476 & \\
 1295 & 16:06:03.27 & +55:24:57.76 & 20.654 & $0.0701 \pm 0.0058$ & $-0.0658 \pm  0.0473$ &  0.78 &  0.090 &  0.428 & \\
 1364 & 16:06:15.43 & +55:24:52.52 & 23.303 & $0.0330 \pm 0.0469$ & $ 0.1256 \pm  0.0582$ &  0.78 &  0.128 &  0.702 & \\
 1475 & 16:06:19.52 & +55:24:30.48 & 23.856 & $0.0490 \pm 0.0149$ & $ 0.1673 \pm  0.0175$ &  0.67 &  0.040 &  0.379 & \\
\enddata
\label{tab:data}
\end{deluxetable}

\end{document}